\documentclass[%
 reprint,
superscriptaddress,
showpacs,preprintnumbers,
twocolumn,
tightenlines,
prb,
]{revtex4-2}

\usepackage{graphicx}
\usepackage{dcolumn}
\usepackage{bm}%
\usepackage{dcolumn}
\usepackage{siunitx}
\usepackage{amssymb}
\usepackage{booktabs}
\usepackage{setspace}
\usepackage{xcolor}
\begin{document}

\title{Single-crystal study of the honeycomb XXZ magnet BaCo$_2$(PO$_4$)$_2$ in magnetic fields}

\author{Xiao Wang}
\affiliation{II.\ Physikalisches Institut, Universit\"at zu K\"oln, Z\"ulpicher Str.\ 77, 50937 K\"oln, Germany}
\author{Rohit Sharma}
\affiliation{II.\ Physikalisches Institut, Universit\"at zu K\"oln, Z\"ulpicher Str.\ 77, 50937 K\"oln, Germany}
\author{Petra Becker}
\affiliation{Abteilung Kristallographie, Institut f\"ur Geologie und Mineralogie, Universit\"at zu K\"oln, Z\"ulpicher Str.\ 49b, 50674 K\"oln, Germany}
\author{Ladislav Bohat\'y}
\affiliation{Abteilung Kristallographie, Institut f\"ur Geologie und Mineralogie, Universit\"at zu K\"oln, Z\"ulpicher Str.\ 49b, 50674 K\"oln, Germany}
\author{Thomas Lorenz}
\affiliation{II.\ Physikalisches Institut, Universit\"at zu K\"oln, Z\"ulpicher Str.\ 77, 50937 K\"oln, Germany}

\date{\today}

\begin{abstract}
We present a study of high-quality BaCo$_2$(PO$_4$)$_2$ single crystals via magnetization, heat-capacity, thermal-expansion and magnetostriction measurements. Sharp anomalies in the thermodynamic properties at $T_N=3.4\,$K reveal a long-range antiferromagnetic order in these single-crystalline samples, which is absent in polycrystalline BaCo$_2$(PO$_4$)$_2$. 
The temperature dependent magnetic susceptibilities for in-plane and out-of-plane magnetic fields are strongly anisotropic and reveal a pronounced easy-plane anisotropy. A Curie-Weiss analysis implies strong orbital magnetism, as it is known from the sister compound BaCo$_2$(AsO$_4$)$_2$ that is discussed as a potential Kitaev spin-liquid material.  
When applying in-plane magnetic fields at low temperature, BaCo$_2$(PO$_4$)$_2$ is driven to another ordered phase at a critical field $\mu_0H_{C1}\approx 0.11\,$T and then undergoes a further field-induced transition to a highly polarized paramagnetic phase at $\mu_0H_{C2}\approx 0.3\,$T, which is again similar to the case of  BaCo$_2$(AsO$_4$)$_2$. In addition, our lowest-temperature data reveal that the field-induced transitions in  BaCo$_2$(PO$_4$)$_2$ become dominated by thermally assisted domain-wall motion.

\end{abstract}

\date{\today}

\maketitle

\section{Introduction}

Studies of the Kitaev model have become a very active research area in condensed matter physics during the past decade, mainly motivated by the expectation to observe novel physical properties such as topological order with exotic excitations and their potential applications for quantum computing \cite{Kitaev2006}. The Kitaev model is formulated for a spin-1/2 honeycomb lattice with bond-directional Ising-type exchange interaction and is exactly solvable with a quantum spin liquid (QSL) ground state. Experimentally, approximate realizations of this model are found in several transition metal compounds with a honeycomb lattice, e.g.  $\alpha$-RuCl$_3$ or $A_2$IrO$_3$ ($A$ = Li, Na), where strong spin-orbit coupling leads to an effective spin-orbit assisted pseudospin-1/2 electronic state for Ru$^{3+}$ (4$d^{5}$) or Ir$^{3+}$ (5$d^{5}$)  ions in the octahedral crystal field \cite{Takayama2015,Jackeli2009,Takagi2019,Khomskii2021,Trebst2022}.
Many fascinating phenomena like fractionalized excitations and quantized thermal Hall effect are reported in these candidate materials, but all of them exhibit a conventional long-range magnetic order at finite temperature under zero magnetic field due to the presence of substantial non-Kitaev interactions \cite{Do2017,Banerjee2017, Kasahara2018}.
Thus, the original Kitaev model has been extended to the generic Heisenberg-Kitaev-$\Gamma$ model with Kitaev $K$, Heisenberg $J$, and symmetry-allowed off-diagonal  $\Gamma$ exchange terms~\cite{Singh2012,Rau2014}.
The non-Kitaev interactions can be sizable in the real materials, which engenders a rich phase diagram and may prompt the corresponding system far away from the ideal Kitaev spin-liquid state \cite{Chaloupka2010,Rau2014}.
Therefore, searching materials with dominant Kitaev interaction is of great importance for the realization of Kitaev QSL.

Recent theoretical proposals demonstrated that 3$d$ transition-metal compounds with Co$^{2+}$ ions might also be a promising platform for the study of Kitaev physics \cite{Sano2018,Liu2018,Liu2020}.
In an octahedral crystal field, the high-spin state of Co$^{2+}$ (3$d^7$) has an electronic configuration of $t^{5}_{2g}e^{2}_{g}$ with total spin $S = 3/2$ and a  fictitious angular momentum $\widetilde{l} = 1$, giving rise to a pseudospin-1/2 doublet in the ground state, akin to the case in Ru$^{3+}$ (4$d^5$) and Ir$^{4+}$ (5$d^5$) \cite{Rau2016,Piwowarska2019,Takagi2019}.
Whereas the bond-directional Kitaev interaction $K$ is always ferromagnetic in the $t^{5}_{2g}e^{2}_{g}$ case, the Heisenberg term $J$ may be reduced by a partial compensation of ferromagnetic and antiferromagnetic contributions resulting from $e_g$-$e_g$, $e_g$-$t_{2g}$, and $t_{2g}$-$t_{2g}$ exchange interactions, which provides an encouraging avenue for the realization of Kitaev QSL \cite{Liu2018,Liu2020}.
As potential realizations, several Co$^{2+}$-based materials with honeycomb lattice have attracted tremendous research interest, such as Na$_2$Co$_2$TeO$_6$, Na$_3$Co$_2$SbO$_6$,  BaCo$_2$(AsO$_4$)$_2$ and BaCo$_2$(PO$_4$)$_2$ \cite{Viciu2007,Wong2016,Lin2021,Yan2019,Wellm2021,Zhong2020,Shi2021,Nair2018}. 
In particular, BaCo$_2$(AsO$_4$)$_2$ has been recently reported to exhibit a non-magnetic ground state under a small magnetic field in the honeycomb $ab$ plane, which may be dominated by Kitaev interaction~\cite{Zhong2020}. 
The magnetism of this material was already studied in the late 1970's and antiferromagnetic ordering at $T_N = 5.4\,$K with a complex magnetic structure was found for single crystals of BaCo$_2$(AsO$_4$)$_2$~\cite{Regnault1977,Regnault1979}.
In contrast, for its sister compound BaCo$_2$(PO$_4$)$_2$ only polycrystalline samples have been available so far, on which only short-range magnetic correlations are  oberved~\cite{Regnault1977,Regnault1979,Nair2018}.
Here, we report a study of  high-quality single crystals of BaCo$_2$(PO$_4$)$_2$, which clearly reveals an antiferromagnetic (AFM) long-range magnetic order with $T_N = 3.4\,$K in zero magnetic field.
Based on magnetization, heat capacity and dilatometer experiments, we derive the magnetic-field temperature phase diagram for in-plane magnetic fields, which reveals many similarities to the case of BaCo$_2$(AsO$_4$)$_2$~\cite{Zhong2020}.

\begin{figure}
	\centering
	\includegraphics[width=8.5cm]{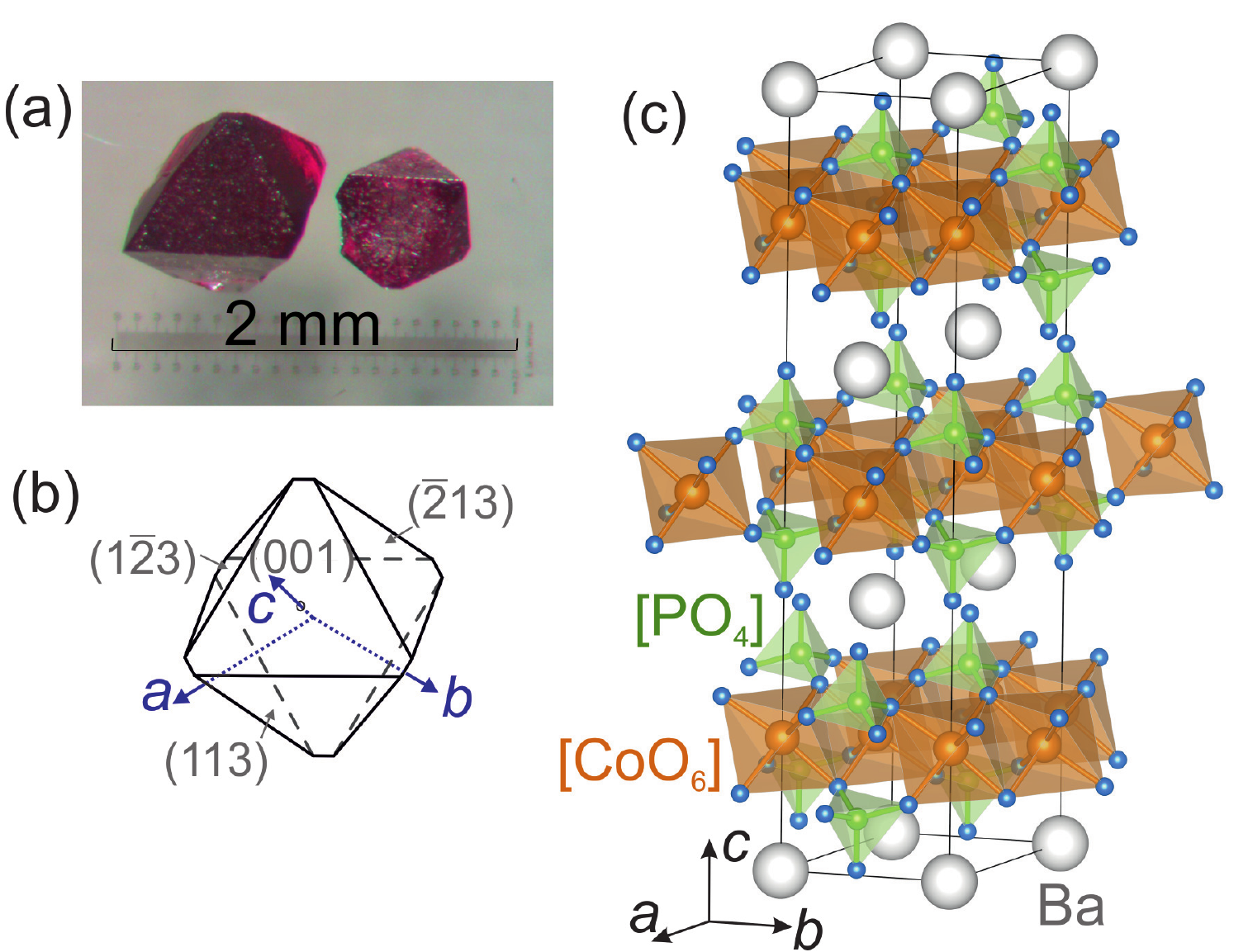}
	\caption{\label{fig:Fig1}
		(a) Hydrothermally grown crystals of BaCo$_2$(PO$_4$)$_2$.
		(b) Typical morphology of grown crystals with pinacoid \{001\} and rhombohedron face (113) plus symmetrically equivalent faces.
		Crystallographic axes are indicated.
		(c) A fraction of the crystal structure of BaCo$_2$(PO$_4$)$_2$ with honeycomb layers of edge-sharing CoO$_6$ octahedra (orange) alternatingly stacked along $c$ with layers of Ba atoms (white) and PO$_4$ groups (green).
		Structure data taken from Ref.~\cite{Bircsak1998}, structure model generated with VESTA~\cite{Momma2011}.
	}
\end{figure}

\section{Experimental}

Crystals of the $\gamma$-modification of BaCo$_2$(PO$_4$)$_2$~\cite{David2013} were grown via a hydrothermal method inspired by the ``guanidinium route"~\cite{Bircsak1998}. 
3.75 mmol BaCO$_3$, 7.5 mmol CoBr$_2$, 7.5 mmol [C(NH$_2$)$_3$]$_2$CO$_3$ and 15 mmol H$_3$PO$_4$, together with 30 ml purified H$_2$O were enclosed in the Teflon liner of a 45 ml autoclave and heated for three weeks.
In a series of experiments $\sim$448\,K turned out to be the optimum temperature for crystal growth.
The grown pink crystals reach dimensions of up to 1 mm$^3$, and their morphology is dominated by the pinacoid \{001\} as well as the rhombohedron \{113\}, see Fig.~\ref{fig:Fig1}.
The trigonal crystals show uniaxial negative optical character and there are no hints to growth defects between crossed polarizers in a microscope.
From XRD data on powdered crystals room temperature lattice constants $a = 4.8296(6)\,$\AA ~and $c = 23.1009(30)\,$\AA\ are refined, in agreement with Ref.~\cite{Bircsak1998}. 
$\gamma$-BaCo$_2$(PO$_4$)$_2$ crystallizes into a trigonal structure with the space group $R\overline{3}$~\cite{Bircsak1998,David2013}, and contains honeycomb layers of slightly distorted edge-sharing CoO$_6$ octahedra, [see Fig.~\ref{fig:Fig1}(c)], which is essential for Kitaev exchange. The nearest intralayer Co-Co distance is $2.80\,$\AA , but there is a large interlayer distance of $7.74\,$\AA\ due to intermediate layers of Ba atoms and tetrahedral [PO$_4$] groups. 

Using commercial setups (Quantum design MPMS or PPMS) the magnetization was measured in magnetic fields up to 1\,T applied either in-plane or out-of-plane for temperatures  from 1.8 to 300\,K, while the low-temperature heat capacity was studied down to 2.5\,K for different in-plane magnetic fields. High-resolution measurements of the relative length changes $\Delta L(T,\mu_0H)/L_0$, either as a function of the temperature or the magnetic field, were obtained on a home-built capacitance dilatometer that was attached to a commercial $^3$He cryostat (Oxford Heliox VL) reaching temperatures below about 300\,mK. Due to the sample geometry, $\Delta L(T,\mu_0H)/L_0$ was measured parallel to the $c$ axis, whereas the magnetic field was again applied in-plane.   

\begin{figure}
	\centering
	\includegraphics[width=8.8cm]{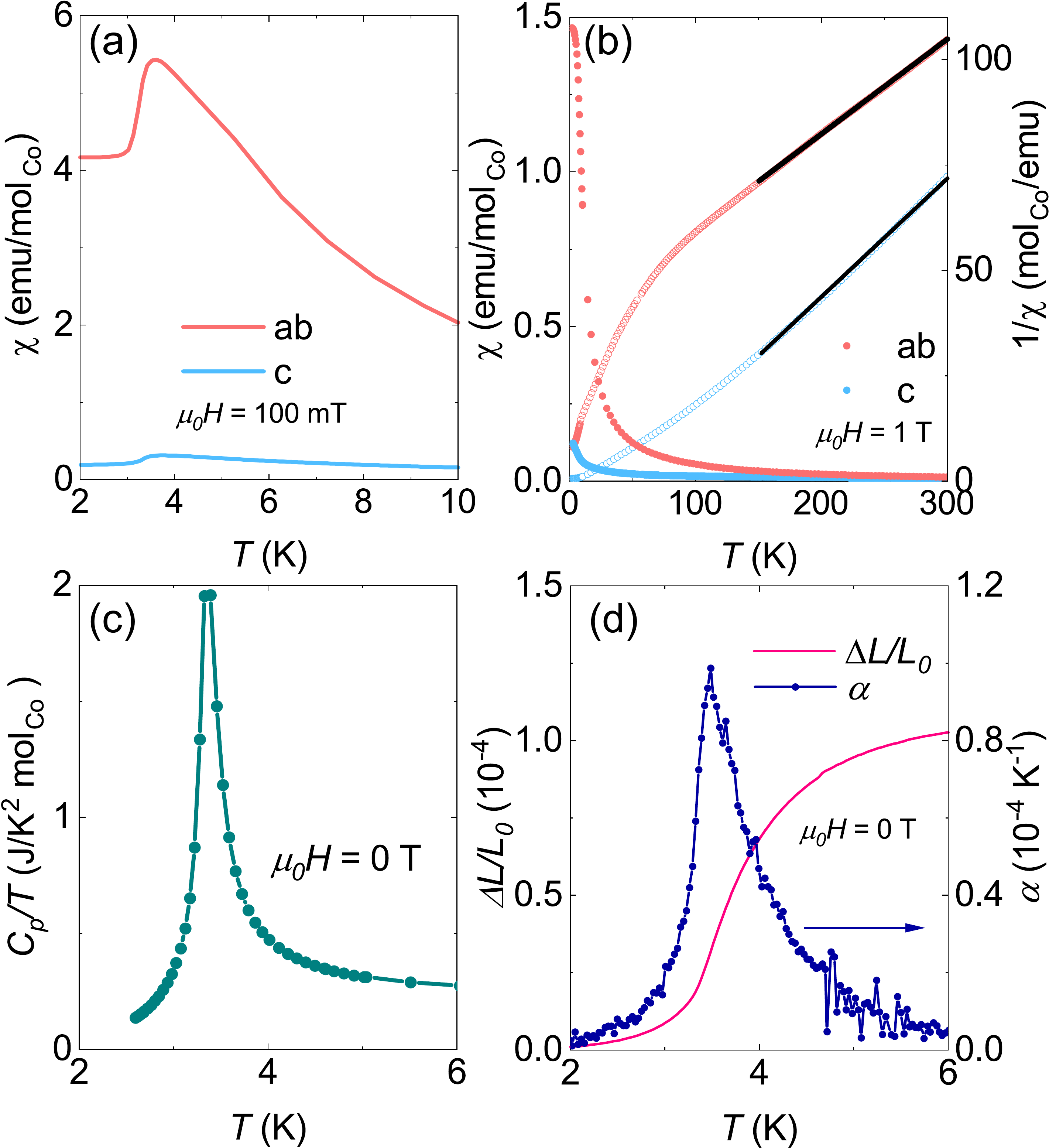}
	\caption{\label{fig:Fig2}
		(a,b) Magnetic susceptibilities $\chi_{ab}$ and $\chi_{c}$ for in-plane and out-of-plane magnetic fields of 0.1 T and 1\,T. Panel (b) also shows  $1/\chi_{i}$ (right axes) together with Curie-Weiss fits (lines) for $T>150\,$K. 
		(c) Zero-field heat capacity and (d) relative length change $\Delta L(T)/L_0$ of the $c$ axis together with the uniaxial thermal expansion coefficient $\alpha_{c} = 1/L_0\,\partial\Delta L(T)/\partial T$.
	}
\end{figure}

\section{Results and discussion}

Figure~\ref{fig:Fig2} illustrates the magnetic phase transition in BaCo$_2$(PO$_4$)$_2$ studied by magnetization, specific heat and thermal expansion.
As all data sets show distinct anomalies, it is clear that there is a sharp transition at $T_{\rm N} = 3.4\,$K. As shown in Fig.~\ref{fig:Fig2}\,(a), the magnetic susceptibilities measured for an in-plane ($\chi_{ab}$) as well as for an out-of-plane ($\chi_{c}$) magnetic field drop below $T_{\rm N}$ implying the phase transition to be antiferromagnetic. In the low-temperature region, $\chi_{ab}$ is more than 10 times larger than $\chi_{c}$, which indicates a strong in-plane anisotropy. 
With increasing temperature this anisotropy continuously decreases to $\chi_{ab}/\chi_{c}\simeq 1.5$ at room temperature.
The $1/\chi_i$ plots in Fig.~\ref{fig:Fig2}\,(b) reveal a strong deviation from a simple Curie-Weiss behavior, but above about 150\,K both curves approach straight lines with similar slopes. Consequently, standard Curie-Weiss (CW) fits restricted to the higher-temperature region yield similar effective magnetic moments $\mu_{eff}^{ab}$ = 5.96 $\mu_B$/Co$^{2+}$ and $\mu_{eff}^c$ = 5.34 $\mu_B$/Co$^{2+}$. These values strongly exceed the spin-only value $g\,\sqrt{S(S+1)}=3.87$ for $S=3/2$ with $g = 2$, but are comparable to other Co$^{2+}$ systems, e.g. BaCo$_2$(AsO$_4$)$_2$, Na$_2$Co$_2$TeO$_6$ and Na$_3$Co$_2$SbO$_6$, signaling large orbital contributions in these materials~\cite{Viciu2007,Zhong2020,Lin2021}.
The Weiss temperatures obtained from the fits for both field directions are strongly different, with $\Theta_{ab} = -166$\,K and $\Theta_{c} = 44$\,K  indicating either predominant antiferromagnetic or ferromagnetic couplings, respectively.
Both, the anisotropic $\Theta_{i}$ as well as the almost isotropic effective magnetic moment are very close to the corresponding values obtained for BaCo$_2$(AsO$_4$)$_2$~\cite{Zhong2020}.
Weiss temperatures of different signs are also observed in the Kitaev candidate $\alpha$-RuCl$_3$, where the anisotropy of the magnetic susceptibility has been suggested to arise from large bond dependent off-diagonal interactions~\cite{Sears2020}.

Figure~\ref{fig:Fig2}\,(c) shows the heat capacity of a BaCo$_2$(PO$_4$)$_2$ single crystal at zero magnetic field which shows one sharp $\lambda$-shaped anomaly at $T_N$, consistent with the magnetic susceptibility results.
It has to be mentioned that there is no sharp transition in polycrystalline samples, instead two broad humps near 3.5 and 6.1\,K are observed~\cite{Regnault1978,Nair2018}.
A similar behaviour is reported for $\alpha$-RuCl$_3$, where in high-quality single crystal samples only one sharp transition is observed at $\sim$8\,K, but for polycrystalline samples or samples with high stacking faults density, the transition peak at $\sim$8\,K becomes weak and broad along with several additional broad  anomalies between 10 and 15\,K~\cite{Cao2016}.
Figure~\ref{fig:Fig2}\,(d) shows the temperature dependent relative length change $\Delta L(T)/L_{0} $ measured along the $c$ axis together with the uniaxial thermal expansion coefficient $\alpha_{c}=1/L_0 \, \partial\Delta L(T)/\partial T$.
Besides confirming the transition at $T_N$, the sharp peak in $\alpha_{c}$ also reveals a pronounced magnetoelastic coupling in BaCo$_2$(PO$_4$)$_2$ because it signals a relative contraction of the interlayer distance by $\Delta L/L_{0}\approx 10^{-4}$ due to the antiferromagnetic order.
Consequently, uniaxial pressure $p_c$ along the $c$ axis will stabilize the ordered phase and a strong initial increase of $\partial T_N/\partial p_i \approx 4\,$K/GPa is estimated via the Clausius-Clapeyron relation $\partial T_N/\partial p_c = V_{mol} T_N \Delta \alpha / \Delta C_p$ with $\Delta \alpha$ and $\Delta C_p$ denoting the peak heights of $\alpha$ and $C_p$, respectively.

\begin{figure}
	\centering
	\includegraphics[width=8.5cm]{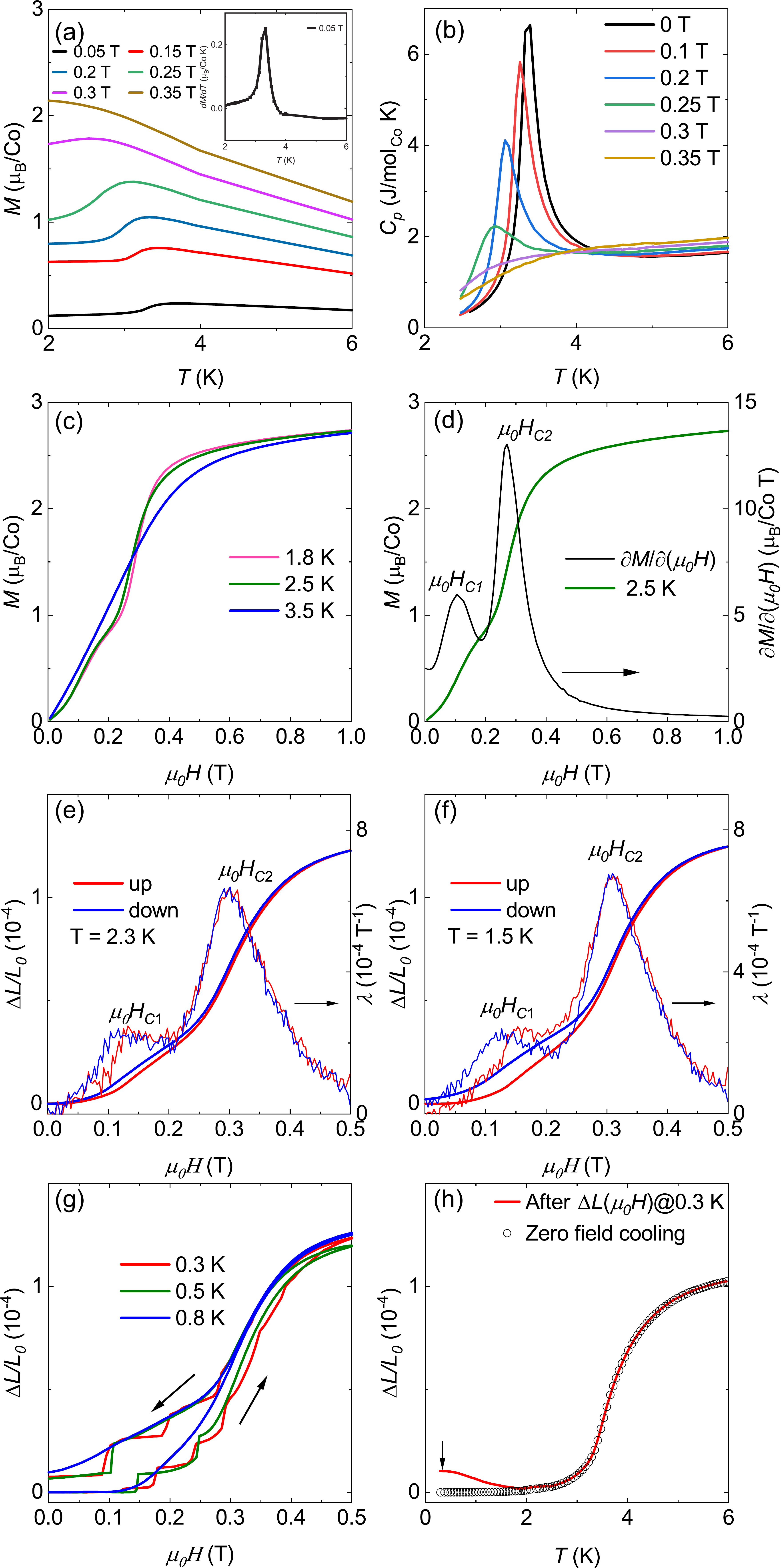}
	\caption{\label{fig:Fig3}
		(a) Magnetization $M(T)$ and (b) heat capacity versus temperature for different in-plane magnetic fields. Inset of panel (a): the first derivative $dM(T)/dT$ of magnetization under magnetic field of 0.05 T. 
		(c) Field dependent $M(\mu_0H)$ for $T= 1.8$ to 3.5\,K, (d) also contains the derivative $\partial M/\partial (\mu_0H)$ with two peaks signaling two critical fields at 0.11\,T and 0.27\,T.
		(e,f) Magnetostriction $\Delta L(\mu_0H)/L_{0}$ of the $c$ axis together with the expansion coefficient $\lambda=1/L_0 \, \partial\Delta L/\partial (\mu_0H)$ at 2.3  and 1.5\,K; the data were obtained with increasing (red) and decreasing (blue) field.  
		(g) $\Delta L(\mu_0H)/L_{0}$ at 0.3, 0.5, and 0.8\,K, where each data set was obtained after zero-field cooling from 6\,K with increasing and decreasing field. Each curve ends with a remnant zero-field expansion.      
		(h) Thermal expansion $\Delta L(T)/L_{0}$ (red line) measured after a magnetostriction cycle at 0.3\,K, which induces a remnant zero-field expansion (marked by the arrow) in comparison to $\Delta L(T)/L_{0}$ ($\circ $) measured after zero-field cooling. 
		 	}
\end{figure}

The influence of in-plane magnetic fields on the phase transition in  BaCo$_2$(PO$_4$)$_2$ is summarized in Fig.~\ref{fig:Fig3}.
As shown in panels~(a,b) the anomalies of the magnetization and of the heat capacity first gradually shift to lower temperature upon increasing the field to 0.2\,T, then the anomalies start to broaden and eventually vanish at about 0.3\,T.
The field dependent magnetization measured at different temperatures is presented in Fig.~\ref{fig:Fig3}\,(c).
Already weak in-plane fields of less than 1\,T are sufficient to magnetize the crystal above $2.5\,\mu_B$/Co$^{2+}$, and below 3\,K this large magnetization is reached via 
two transitions with critical fields $\mu_0H_{C1}$ and $\mu_0H_{C2}$ defined via the maxima of $\partial M/\,\partial (\mu_0H)$ as is exemplarily shown in panel~(d).
These magnetic field induced transitions are also observed in the magnetostriction data $\Delta L(\mu_0H)/L_{0}$ measured along the out-of-plane direction with the magnetic field being applied in-plane as shown in Fig.~\ref{fig:Fig3}\,(e).
The magnetostriction data were measured upon increasing and decreasing the magnetic field, but for $T=2.3\,$K there is only little hysteresis around $H_{C1}$, as is seen in both $\Delta L(\mu_0H)/L_{0}$ and its derivative $\lambda=1/L_0 \, \partial\Delta L/\partial (\mu_0H)$. This hysteresis is somewhat enlarged at $T=1.5\,$K, see panel~(f), while performing a field cycle at 800\,mK not only further enlarges the hysteresis, but also results in a significant remnant zero-field expansion.
At even lower temperature, the $\Delta L(\mu_0H)/L_{0}$ curves change their character to a sequence of more or less regular plateaus with intermediate jumps, see panel~(g).
Note that before each field cycle the crystal has been heated to well above $T_N$, typically to 6\,K, and then cooled down in zero field. 

The discontinuous low-temperature magnetostriction curves resemble so-called Barkhausen jumps of ferromagnets, which are typically observed when the magnetization reversal is dominated by the domain-wall motion between differently oriented domains. Microscopically, the plateau regions result from the pinning of domain walls until the driving force on the domain walls due to the continuously varying external field overcomes the pinning force and causes an avalanche-like motion to another strong-enough pinning center. Because thermal fluctuations facilitate the motion of domain walls, the observed temperature dependence of the low-temperature magnetostriction curves can be naturally ascribed to a temperature-assisted domain-wall motion in an external driving field. This conclusion is further supported by the zero-field thermal expansion curves $\Delta L(T)/L_{0}$ displayed in Figure~\ref{fig:Fig3}\,(h). The curve shown by open symbols was measured upon heating directly after zero-field cooling and thus corresponds to thermal equilibrium. In contrast, the result given by the red line has been recorded after the magnetostriction measurement at 300\,mK shown in panel~(g) and thus starts from the remnant zero-field expansion, which thermally decays upon increasing temperature. Consequently, this $\Delta L(T)/L_{0}$ curve, displayed as a red line in panel~(h), decreases with increasing temperature until it merges with the thermal-equilibrium curve at about 2\,K. 

\begin{figure}
	\centering
	\includegraphics[width=9.5cm]{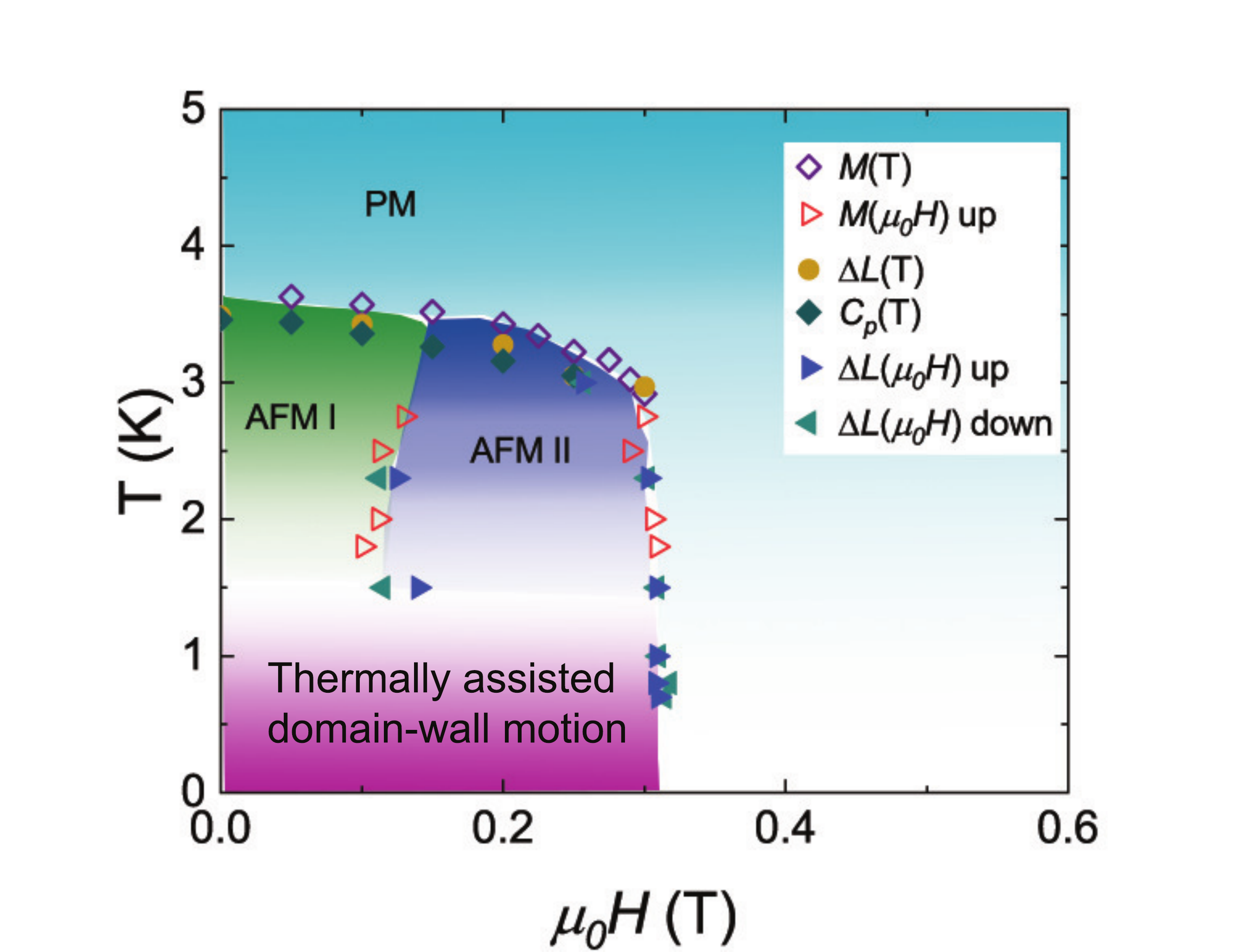}
	\caption{\label{fig:Fig4}
		Phase diagram of BaCo$_2$(PO$_4$)$_2$ for in-plane magnetic fields obtained from heat capacity ($C_p$), uniaxial expansion ($\Delta L$) and magnetization (M) measurements. Transition temperatures and fields are defined by the peak positions of $C_p(T)$ and of the first derivatives of $\Delta L(T,\mu_0H)$ or $M(T,\mu_0H)$ with respect to $T$ or $\mu_0H$, respectively.   
	}
\end{figure}

By combining the critical temperatures and fields obtained above we construct the temperature versus in-plane magnetic field phase diagram in Fig.~\ref{fig:Fig4}.
As a function of temperature there are sharp transitions from the paramagnetic phase (PM) to either an antiferromagnetic phase AFM~I or AFM~II. The transition temperature only weakly decreases with increasing field up to about 300\,mT, but then very rapidly drops to zero.
These sharp transitions signal the evolution of long-range magnetic order in our single-crystalline samples of BaCo$_2$(PO$_4$)$_2$, in sharp contrast to previous heat-capacity and neutron data obtained on polycrystalline samples, which revealed short-range correlations and/or a mixture of helical and collinear phases~\cite{Regnault1978,Regnault2018,Nair2018}. Instead, our phase diagram of BaCo$_2$(PO$_4$)$_2$ strongly resembles the phase diagrams obtained on single crystals of BaCo$_2$(AsO$_4$)$_2$, with the main difference that the transition temperatures and fields of BaCo$_2$(PO$_4$)$_2$ are reduced by about 40\,\%. Thus, one may expect similar magnetic structures of single-crystalline samples of both materials. 
In analogy to $\alpha$-RuCl$_{3}$, the potential occurrence of a Kitaev spin liquid phase was suspected for BaCo$_2$(AsO$_4$)$_2$ in the intermediate field range where magnetic order is suppressed, but the completely polarized state is not yet reached~\cite{Zhong2020}, and a recent terahertz spectroscopy study on BaCo$_2$(AsO$_4$)$_2$ gave some evidence for the existence of a very narrow intermediate phase region ($\sim 0.05\,$T) with continuous excitations directly above the in-plane $H_{c2}$. Because our macroscopic data of BaCo$_2$(PO$_4$)$_2$ can neither support nor contradict this conjecture, analogous spectroscopic studies on the BaCo$_2$(PO$_4$)$_2$ single crystals would be very intriguing.

According to the spherical polarization neutron analysis data, the in-plane zero-field magnetic structure (AFM\,I) of BaCo$_2$(AsO$_4$)$_2$ is characterized by a stacking of quasi-ferromagnetic zig-zag chains running along the $b$ axis~\cite{Regnault1990}.
These spins also slightly cant from $b$ axis and follow the sequence of up-up-down-down, and could be easily rotated at very low energy cost due to the weakly effective inter-chain couplings~\cite{Regnault1990}. 
A small magnetic field (0.25\,T for BaCo$_2$(AsO$_4$)$_2$) is strong enough to drive the magnetic structure to the AFM\,II phase, of which the in-plane propagation vector is locked into a commensurate wavevector $k_{C}=(1/3, 0)$ and its magnetization is close to 1/3 of the saturation value, implying an up-up-down arrangement of the almost ferromagnetic chains under magnetic fields~\cite{Regnault1977,Regnault1990}.
Along the stacking axis $c$, very limited coherence lengths are reported for both ordered phases of BaCo$_2$(AsO$_4$)$_{2}$, which naturally can be traced back to the quasi-2D magnetic character with very weak inter-layer coupling and, in addition, the broken symmetries of the paramagnetic phase also allows the formation of different magnetic domains, such as configuration domains and 180$^{\circ}$ domains~\cite{Regnault1990,Brown2006}.

Overall, our results obtained on BaCo$_2$(PO$_4$)$_2$ single crystals are remarkably similar to those of BaCo$_2$(AsO$_4$)$_2$.
Apart from the rather identical general shape of the phase diagrams of both materials, 
the magnetization of BaCo$_2$(PO$_4$)$_2$ in the intermediate phase AFM\,II is also close to 1/3 of its saturation value, see Fig.~\ref{fig:Fig3}\,(c,d), and both materials show an anomalous hysteresis behavior~\cite{Regnault1990}.
For BaCo$_2$(PO$_4$)$_2$,  the field dependent $\Delta L(\mu_0H)/L_{0}$ curves below 2\,K become systematically different for the field-increasing and the field-decreasing run, see Fig.~\ref{fig:Fig3}\,(e,f,g), and a qualitatively similar behavior has been reported for the magnetic-field and temperature-dependent propagation vector of BaCo$_2$(AsO$_4$)$_2$~\cite{Regnault1990,Regnault2018}.
After zero-field cooling below about 2/3 of the zero-field $T_{N}$, the propagation vector shows an incommensurate-commensurate transition as a function of increasing field, which is only partly reversed or even remains absent in the subsequent field-decreasing run.
Finally, our lowest-temperature $\Delta L(\mu_0H)/L_{0}$ curves of Fig.~\ref{fig:Fig3}\,(g) indicate thermally assisted domain-wall motion, which could result from the partial coexistence of the phases AFM\,I, AFM\,II with fully polarized regions and/or from the field-dependent re-population of different domains~\cite{Brown2006}.
We are not aware of similar observations in BaCo$_2$(AsO$_4$)$_2$, but apparently there are hardly any studies of this material in the temperature range below 1.5\,K.

Our experimental results cannot identify the spin structure of BaCo$_2$(PO$_4$)$_2$, but the characterization of the high quality single crystals have demonstrated close similarities of the physical properties between BaCo$_2$(PO$_4$)$_2$  and its sister compound BaCo$_2$(AsO$_4$)$_2$. 
Very recent neutron scattering measurements on the BaCo$_2$(AsO$_4$)$_2$ single crystals have shown that the frustrated $XXZ$-J$_{1}$-J$_{3}$ model better describes the measured inelastic structure factor rather than the Kitaev JK$\Gamma$$\Gamma^{'}$ model~\cite{Halloran2022}. These results do not support a dominant Kitaev-type interaction in the BaCo$_2$(AsO$_4$)$_2$, but as explicitly pointed out, this material may nonetheless be close to a spin-liquid state which results from geometrical frustration and the pronounced two-dimensional nature of its magnetism. In fact, continuous excitations are observed in BaCo$_2$(AsO$_4$)$_2$ by  terahertz spectroscopy as well as by inelastic neutron experiments~\cite{Halloran2022,Zhang2022}. As such a continuum is a solid evidence of the emergence of spin-liquid behavior, our results on BaCo$_2$(PO$_4$)$_2$ single crystals suggest the occurrence of similar continuous excitation spectra in this sister compound.  This issue asks for further clarification by terahertz spectroscopy or neutron diffraction studies on single crystal samples of BaCo$_2$(PO$_4$)$_2$.

\section{Summary}

We have successfully synthesized high-quality single crystals of BaCo$_2$(PO$_4$)$_2$ on which we observe a strong easy-plane magnetic anisotropy and a long-range antiferromagnetic magnetic order at $T_N=3.4\,$K, whereas in polycrystalline samples of this material only short correlations are observed. 
For in-plane magnetic fields we derive a phase diagram with 2 antiferromagnetically odered phases in close vicinity to the almost fully magnetized state, which strongly resembles the phase diagram of the isostructural BaCo$_2$(AsO$_4$)$_2$. This latter material has been discussed as a 
promising candidate to realize a Kitaev spin liquid~\cite{Liu2018,Liu2020,Zhong2020}. However, the $XXZ$-J$_{1}$-J$_{2}$-J$_{3}$ model yields an alternative description of the magnetism of these layered materials and even without dominant Kitaev-type interaction the close vicinity of competing phases might give rise to another kind of 
quantum spin liquid behavior~\cite{Das2021,Halloran2022,Zhang2022,Maksimov2022new}. In order to clarify these issues, additional investigations on related model materials appear promising and we hope that these findings on single-crystalline samples of BaCo$_2$(PO$_4$)$_2$ will stimulate further experiments on this interesting class of materials.

\begin{acknowledgments}
We thank M.~Braden, A.~Bertin, C.~Hickey, and S.~Trebst for stimulating discussions. We acknowledge support by the German Research Foundation via Project No.277146847-CRC1238 (Subprojects A02 and B01).
\end{acknowledgments}



\providecommand{\noopsort}[1]{}\providecommand{\singleletter}[1]{#1}%

\end{document}